# Observation of low-frequency Raman peak in layered WTe₂

Hirofumi Nema[1*], Yasuhiro Fujii1, Eiichi Oishi[2] and Akitoshi Koreeda[1]

[1]*Department of Physical Sciences, Ritsumeikan University, 1-1-1 Nojihigashi, Kusatsu, Shiga 525-8577, Japan*
[2]*Research Organization of Science and Technology, Ritsumeikan University, 1-1-1 Nojihigashi, Kusatsu, Shiga 525-8577, Japan*

E-mail: h-nema@fc.ritsumei.ac.jp

WTe₂ recently attracted considerable attention as a layered material exhibiting ferroelectricity, giant magnetoresistance, and pressure-induced superconductivity. In this study, we performed Raman spectroscopy on bulk WTe₂, including the unreported low-frequency region. A novel Raman peak (P0) was found at approximately 9 cm$^{-1}$ in addition to the seven already known peaks (P1–P7). Furthermore, the angular and polarization dependence of the spectra revealed that the novel peak had A₁ symmetry. The existence of this novel peak is consistent with first-principles calculations by another group. Our work paves the way for studying the low-frequency vibration modes of atomic-layer ferrolectric films.

## 1. Introduction

Transition-metal dichalcogenides (TMDs) have attracted considerable interest as potential layered materials for advanced devices. TMDs are potentially promising in optoelectronic device applications. This is because some TMDs realize a single-atomic-layer film with a direct band-gap from the indirect semiconducting bulk.[1, 2] Secondly, TMDs are potential materials for valley electronics. The achievement of complete dynamic valley polarization was confirmed in TMDs.[3–5] Thirdly, TMDs exhibit potential as excitonic devices. In TMDs, excitons with high binding energies are realized because of the confinement andweak shielding effects.[6–10] Such excitons are stable at room temperature[11] and can travel long distances, thus, providing application advantages.

Among the many TMDs, $WTe_2$ exhibits remarkable electronic, magnetic, and dielectric properties. The electronic properties were investigated under high pressure using diamond anvil cell, and pressure-induced superconductivity was determined.[12,13] Regarding magnetic properties, this layered material exhibits a large unsaturated magnetoresistance despite being a non-magnetic compound.[14–16] Moreover, ferroelectric properties have been observed near room temperature in two or three atomic-layer films[17, 18] and bulk.[19]

In addition to the studies mentioned above, $WTe_2$ has also been studied using Raman spectroscopy.[20] According to the study, group theory predicts 33 Raman-active modes ($11A_1+6A_2+5B_1+11B_2$) for bulk$WTe_2$ in the Brillouin zone center. Of these possible modes, five $A_1$ and two $A_2$ modes were successfully observed. However, they did not investigate the Raman modes in the low-frequency region, approximately below 50 cm$^{-1}$. Such investigations are important because interesting low-frequency vibration modes sometimes appear in layered materials.[23–41] In this study, we performed Raman spectroscopy of bulk $WTe_2$ down to approximately 5 cm$^{-1}$ and observed unreported novel vibration modes. Furthermore, the symmetry of the mode was clarified through detailed measurements of the polarization-dependent and angle-resolved spectra.

## 2. Methods

We employed a commercially available$WTe_2$ crystal[42] (space group $Pmn2_1$, point group$C_{2v}$) grown by the flux zone method for Raman scattering measurements. The measurements were conducted under ambient conditions using a 532 nm excitation source (Nd: YAG laser, Oxxius LCX-532S-300) in backscattering geometry, as shown in Fig. 1(a). The crystallographic abplane of our sample was aligned parallel to the xy-plane in the figure. The irradiation direction of the excitation light (~4 mW) was parallel to the z direction. We used notch filters[43] to suppress the elastically scattered light in the band of approximately $\pm$ 5 cm$^{-1}$ entering the spectrometer. The filtered light was collected into a single spectrometer (Jobin Yvon, HR320) coupled to a CCD detector. The measurement of circular polarization (helicity) dependence was performed by inserting a quarter-wave plate, as illustrated in Fig. 1(b). For angle-resolved measurements, a half-wave plate was positioned ahead of the objective lens (Fig. 1(c)) and rotated it to acquire data following the similar approach as reported elsewhere.[44]

## 3. Results and discussion

The Raman spectrum of the bulk samples are shown in Fig. 2. In the spectrum, seven peaks (P1–P7) appeared in the region higher than 50 cm$^{-1}$. These peaks were observed at 79.1 cm$^{-1}$ (P1), 88.9 cm$^{-1}$ (P2), 110.2 cm$^{-1}$ (P3), 115.3 cm$^{-1}$ (P4), 131.4 cm$^{-1}$ (P5), 161.8 cm$^{-1}$ (P6), and 208.9 cm$^{-1}$ (P7). The positions of these peaks are almost the same as those reported by Kong et al.[20] This indicates that our sample quality is similar to theirs. In this study, we successfully performed low-frequency Raman spectroscopy and found a novel peak (P0) approximately 9 cm$^{-1}$, in addition to the seven known peaks (P1–P7).

To reveal the nature of the peak P0 in more detail, we investigated the polarization dependence of the spectra. Figure 3(a) shows the linear polarization dependence obtained by adjusting the in-plane direction of the sample crystal. The spectrum measured in parallel nicols (HH) is indicated by the magenta curve. In this curve, P1 and P4–P7 are observed. In crossed nicols (HV) configuration, we observed a spectrum represented blue curves. In contrast to the HH spectrum, peaks P2 and P3 appeared with noticeable intensity, whereas the others disappeared. Figure 3(b) shows the circular polarization dependence of Raman spectra. The blue curve indicates the spectrum captured under the LR configuration, where L (R) represents left (right) circular polarization. All eight peaks appear on this spectrum. The LL spectrum obtained after changing the polarization configuration is plotted as a red curve. We confirmed the appearance of peaks P0, P1, and P4–P7. The polarization-dependent results are summarized in Table I.

The Raman intensity ($I$) formula[45, 46] $I \propto |\langle \epsilon_o | R_j | \epsilon_i \rangle|^2$ is useful for understanding the obtained polarization dependence. Here, $|\epsilon_i\rangle$, $|\epsilon_o\rangle$, and $R_j$ denote the polarization vectors of the incident light, the scattered light, and the Raman tensor, respectively. We represent the linearly polarized wave vectors as $|H\rangle = (1, \quad 0, \quad 0)^T$ and $|V\rangle = (0, \quad 1, \quad 0)^T$. As in the previous study,[47] the left and right circularly polarized wave vectors are expressed as $|L\rangle = \frac{1}{\sqrt{2}}(1, \quad i, \quad 0)^T$, $|R\rangle = \frac{1}{\sqrt{2}}(1, \quad -i, \quad 0)^T$. The Raman tensors for $A_1$, $A_2$, $B_1$, and $B_2$ are expressed as follows[20]

$$A_1 = \begin{pmatrix} a & 0 & 0 \\ 0 & b & 0 \\ 0 & 0 & c \end{pmatrix}, A_2 = \begin{pmatrix} 0 & d & 0 \\ d & 0 & 0 \\ 0 & 0 & 0 \end{pmatrix}, B_1 = \begin{pmatrix} 0 & 0 & e \\ 0 & 0 & 0 \\ e & 0 & 0 \end{pmatrix}, B_2 = \begin{pmatrix} 0 & 0 & 0 \\ 0 & 0 & f \\ 0 & f & 0 \end{pmatrix}.$$

The calculation results of $|\langle \epsilon_o | R_j | \epsilon_i \rangle|^2$ obtained using these vectors and tensors are listed in Table II. As shown in the table, $|\langle \epsilon_o | R_j | \epsilon_i \rangle|^2$ for the $A_1$ ($A_2$) mode becomes zero in HV configuration (HH and LL configurations). These configuration dependencies completely explain our experimental results for the peaks P1–P7. For the $B_1$ and $B_2$ modes, $|\langle \epsilon_o | R_j | \epsilon_i \rangle|^2$ is always zero, regardless of the polarization configuration. Therefore, we could not observe these two modes in the current experimental setup. Because our experimental novel peak P0 disappears only in the HV polarization configuration, the peak is identified as an $A_1$ symmetry mode.

We verified the vibrational symmetry of peak P0 from different perspectives. Figure 4 shows the in-plane polarization angle dependence of the observed Raman peak intensities for HH and HV configurations. In this figure, $\theta$ represents the angle between the sample crystal and the polarization direction of the irradiated laser light (inset in Fig. 4). As can be seen, the intensities showed oscillatory behavior. In the HH configuration, P0 and P4–P7 (P2 and P3) peaks exhibit two-fold (four-fold) symmetry. In the HV configuration, a four-fold symmetry was confirmed for all peaks. This observed symmetry is consistent with the experimental results of other group,[20] except for the unreported peaks.

To interpret the oscillatory behaviors, we considered the angle dependence of Raman intensity using the formula $I \propto |\langle \epsilon_o | R_j | \epsilon_i \rangle|^2$. Table III shows that the intensities can be easily obtained by assuming that the polarization vector of the light is $|H\rangle = (\cos\theta, \sin\theta, 0)^T$ or $|V\rangle = (-\sin\theta, \cos\theta, 0)^T$. As shown in the table, the intensity of the $A_1$ modes is expected to exhibit two-fold (HH configuration) and four-fold symmetries (HV configuration). By contrast, the expected behavior of the intensity of the $A_2$ modes was only two-fold symmetry. From these expected behaviors, we can interpret the experimentally observed peaks P0, P1, and P4–P7 to correspond to the $A_1$ mode and P2 and P3 to the $A_2$ mode, respectively. The experimental results of the angle dependence also support the peak P0 corresponding to the $A_1$ mode. The obtained information on Raman tensor elements from the experimental results are presented in the appendix.

First principles calculations of the vibration modes of bulk WTe$_2$ were reported in a previous study by Kong et al.[20] Their calculations indicated the possible existence of a low-frequency Raman active vibration mode ($8.9$ cm$^{-1}$, $A_1$) below $10$ cm$^{-1}$. The calculated peak position was close to that of P0 ($8.6$ cm$^{-1}$, $A_1$). Additionally, the calculated positions of the other seven peaks were also close to our experimental values, as shown in Table IV. Thus, our results do not significantly conflict with their calculations.

The observed novel peak P0 may be attributed to the inter-layer vibration mode because it is located in the low frequency region. In TMDs, the weak van der Waals interaction between layers acts as a restoring force for the inter-layer vibration. Therefore, the frequencies of the inter-layer vibrations can be low. Such vibration modes have been observed in many types of TMDs below approximately $50$ cm$^{-1}$. As inter-layer vibration modes, the shear ($E_{2g}$) and layer breathing ($A_{1g}$) modes are well-recognized. The former corresponds to the in-plane relative vibrations of the atomic layers, whereas the latter corresponds to out-of-plane vibrations. The $A_1$ symmetry of the observed peak P0 can be attributed to the layer-breathing mode.

To determine whether peak P0 is truly attributed to interlayer vibration mode, investigating the layer-number dependence of the peak position is critical and necessary. Thus far, linear chain models (LCM) have adequately described such layer-number dependencies.[48] This model considers one atomic layer as one weight and the interaction between the atomic layers as a spring connecting the weights. In addition, the interactions between adjacent layers were considered the most significant. If we measured the layer-number dependence of the P0 peak position and explained the measured dependence using the LCM, we could regard the peak as an inter-layer vibration mode. However, these measurements are difficult because ambient air rapidly oxidizes atomic-layer samples.[49] In the near future, it is desirable to realize measurements that can overcome the difficulty of oxidation.

## 4. Summary

In summary, we conducted Raman scattering measurements on bulk $WTe_2$ including in the low frequency region. A novel Raman peak ($A_1$) was captured at approximately 9 $cm^{-1}$ in addition to the seven known peaks ($5A_1 + 2A_2$). The polarization and angular dependence of the novel peak are consistent with the point group ($C_{2v}$) analysis and previously reported first-principles calculations. To deepen our understanding of the low-frequency peaks, measurements of atomic-layer films are desirable.

## Acknowledgment

H. Nema thanks Mr. S. Saito for technical support regarding Raman spectroscopy. This work was supported in part by JSPS KAKENHI (Grant Numbers JP19H05618, JP21H01018) and the Kyoto University Nanotechnology Hub in the "Advanced Research Infrastructure for Materials and Nanotechnology Project" sponsored by the Ministry of Education, Culture, Sports, Science and Technology (MEXT), Japan.

## Appendix: Information on Raman tensor elements (Peak P0)

Through the analysis of experimental data on the angle dependence of Raman intensities, we can obtain the information regarding the the Raman tensor elements. In such analysis, the Raman tensors with complex elements are necessary.[50)] In the case of $A_1$ mode, the tensor is expressed as follows

$$A_1 = \begin{pmatrix} a & 0 & 0 \\ 0 & b & 0 \\ 0 & 0 & c \end{pmatrix} = \begin{pmatrix} ae^{i\varphi_a} & 0 & 0 \\ 0 & be^{i\varphi_b} & 0 \\ 0 & 0 & ce^{i\varphi_c} \end{pmatrix}.$$

Considering this tensor, the following equations for the angle dependence of Raman intensity is derived.

$$I^{\parallel} \propto |a|^2\cos^4\theta + |b|^2\sin^4\theta + 2|a||b|\cos\varphi \sin^2\theta \cos^2\theta. \quad (A\cdot1)$$

$$I^{\perp} \propto \frac{|a|^2 - 2|a||b|\cos\varphi + |b|^2}{4}\sin^2\theta. \quad (A\cdot2)$$

Here $I^{\parallel}$ ($I^{\perp}$) and $\varphi$ indicate the Raman intensity for the HH (HV) configuration and a relative phase difference between two tensor elements ($\varphi=\varphi_b-\varphi_a$). The results of data fitting using the equation (A·1) and (A·2) are shown in Figure A·1 (a) and (b). The obtained fitting parameters are listed in Table A·1.

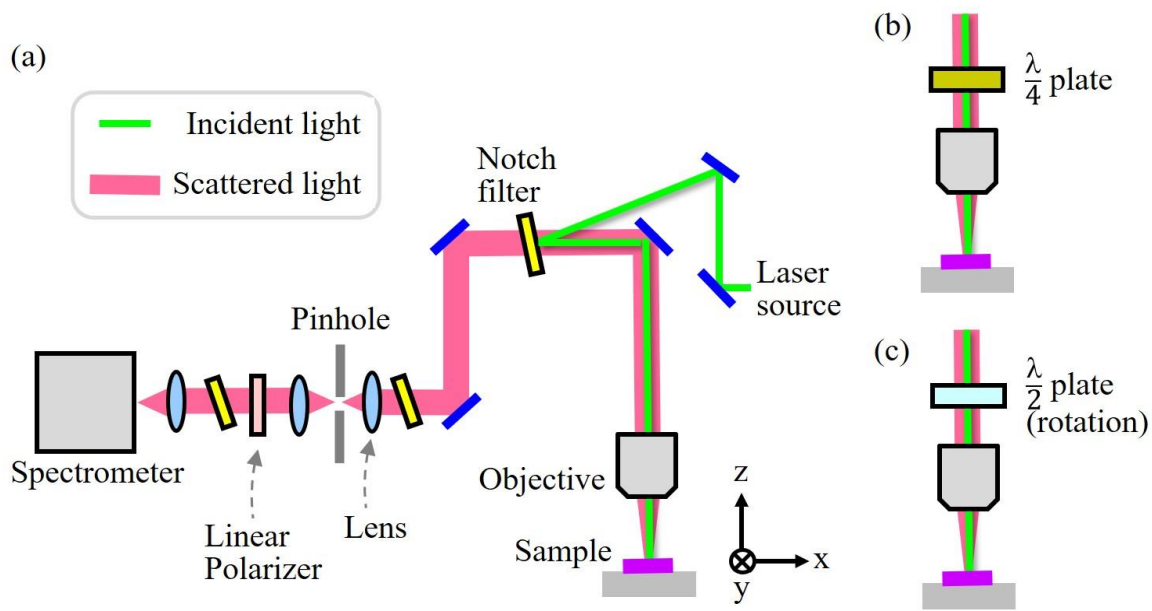

**Fig. 1.** (a) Experimental system for the Raman spectroscopy. The setups for (b) helicity-resolved and (c) angle-resolved measurements.

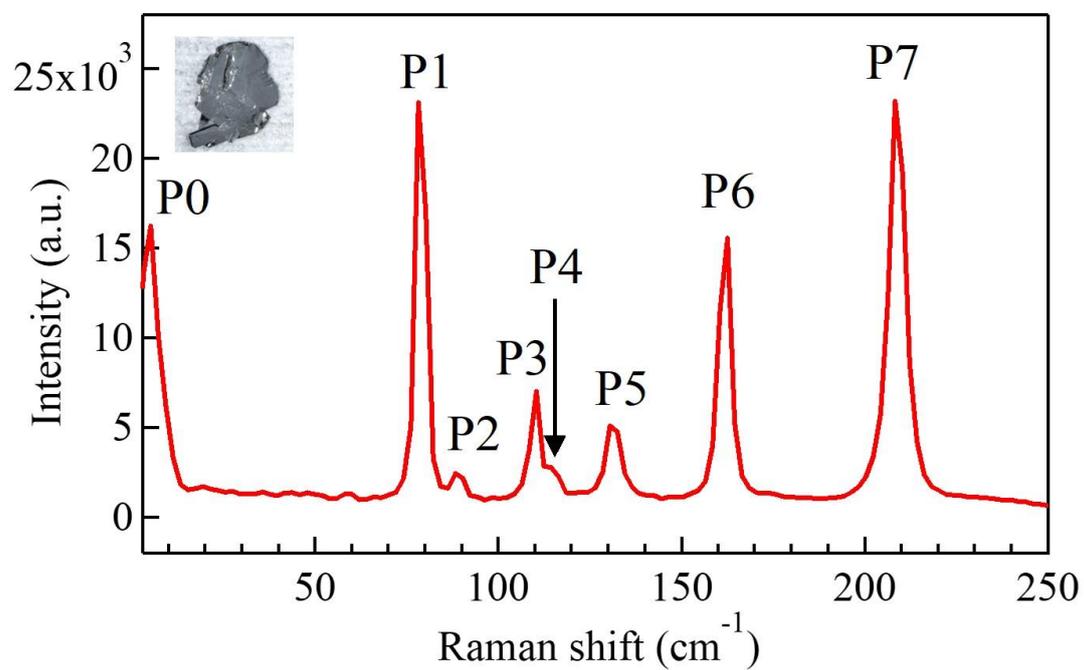

**Fig. 2.** Raman spectrum in bulk WTe₂. The inset shows the picture of our sample.

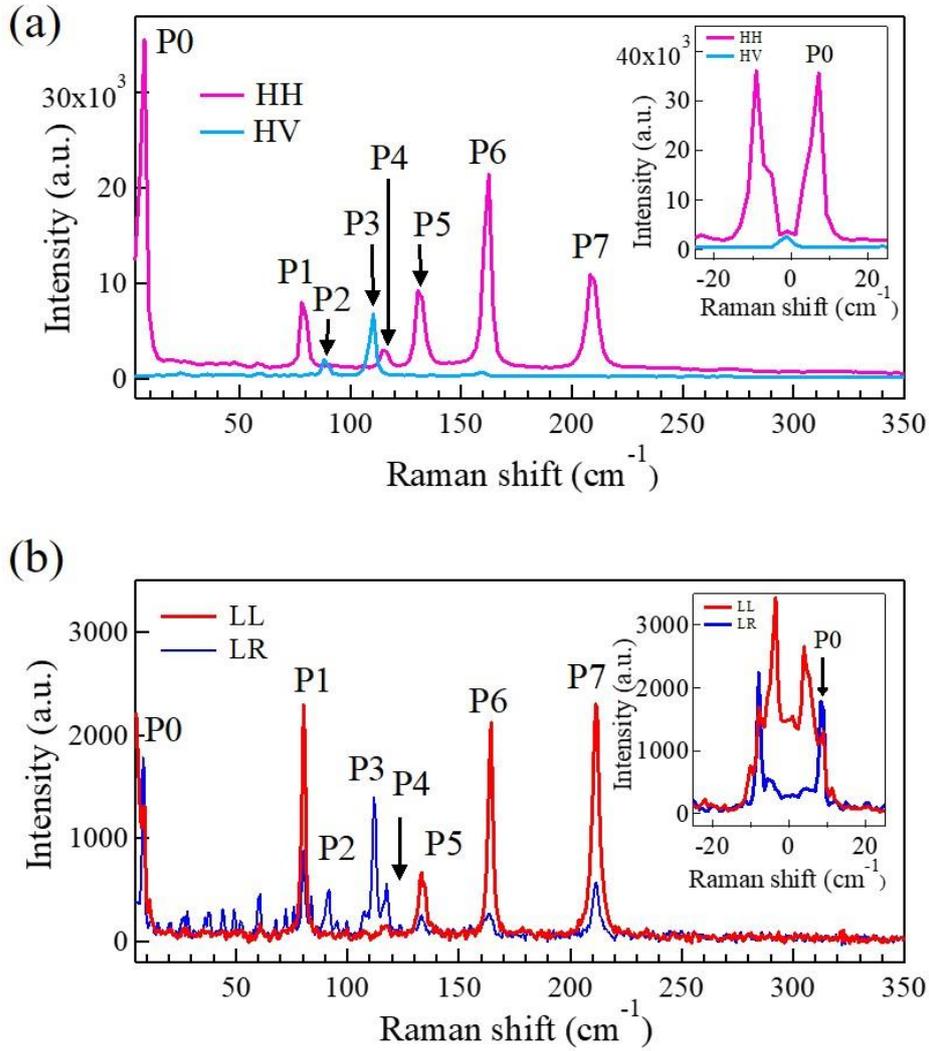

**Fig. 3.** Polarization-dependent Raman spectra in WTe$_2$. (a) Linear polarization dependence, (b) Circular polarization dependence (helicity) dependence. The insets include the anti-Stokes region.

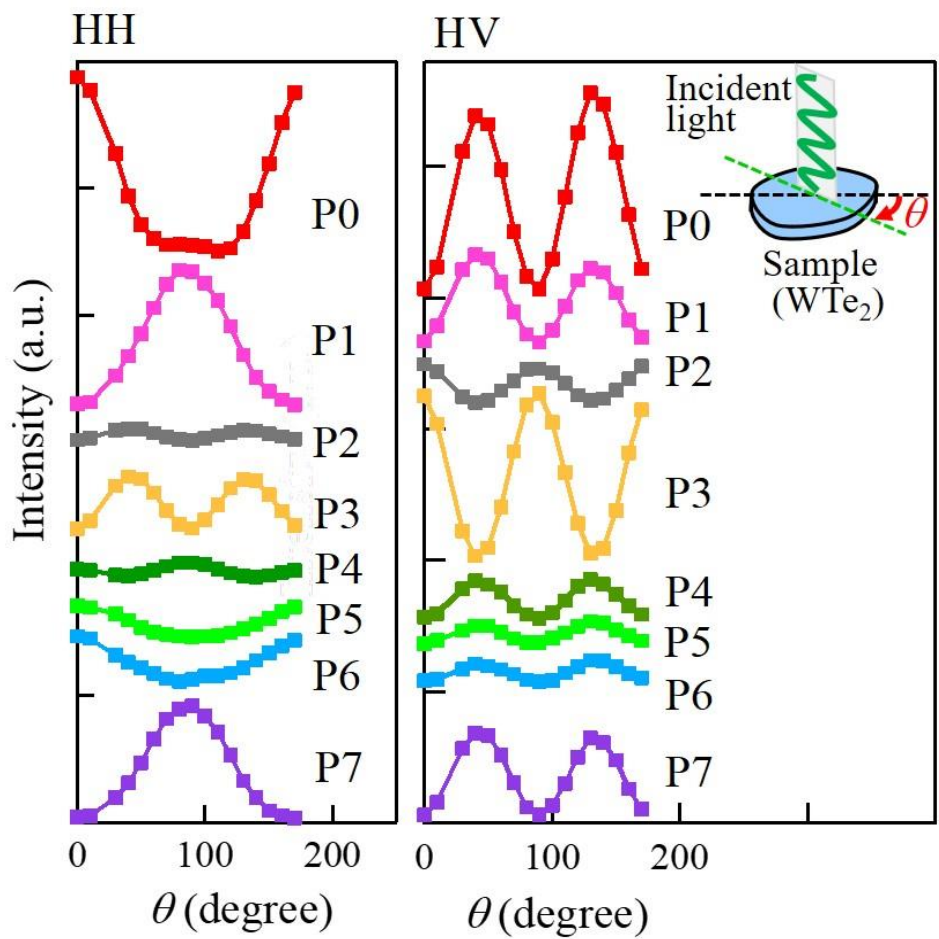

**Fig. 4.** Angular dependence of the intensity in eight Raman peaks (P0–P7) observed in bulk WTe$_2$. Each plot is properly offset for visibility. Inset shows a schematic of angle-dependent measurements.

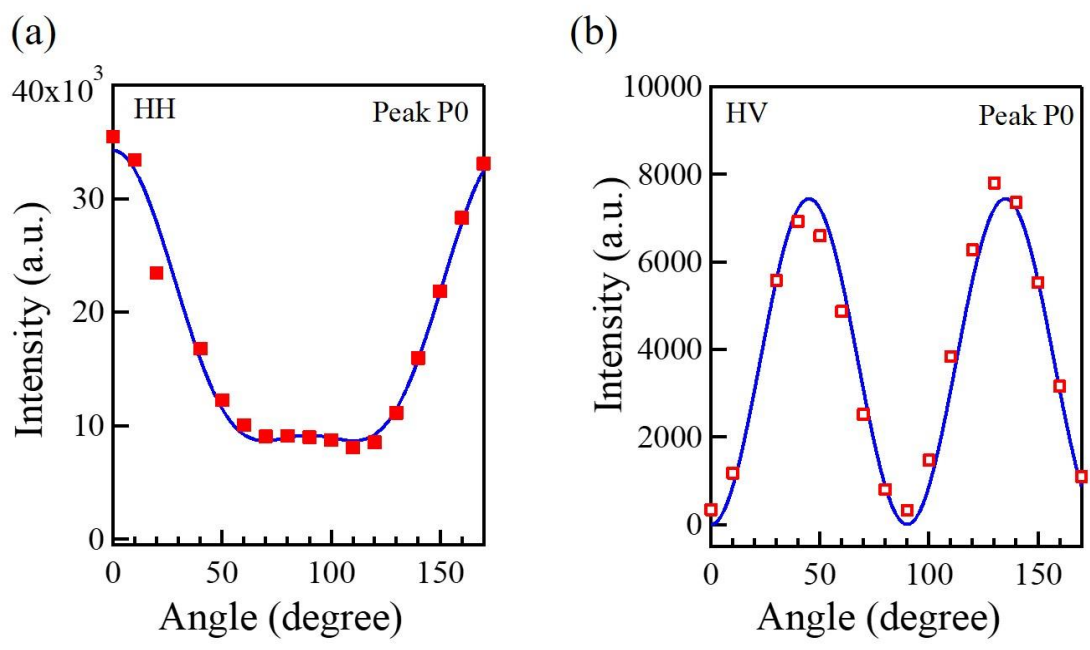

**Fig. A·1.** Angular dependence of the Raman intensity (peak P0) in the (a) HH and (b) HV configuration. The blue solid curves indicate the fitting results.

**Table I.** Summary of experimental polarization dependences. The presence (absence) of Raman peaks in the spectra is represented by the symbol ○ (×). The symmetry of the vibration modes assumed from the experimental results is given in the bottom row of the table.

| | P0 | P1 | P2 | P3 | P4 | P5 | P6 | P7 |
|---|---|---|---|---|---|---|---|---|
| HH | ○ | ○ | × | × | ○ | ○ | ○ | ○ |
| HV | × | × | ○ | ○ | × | × | × | × |
| LL | ○ | ○ | × | × | ○ | ○ | ○ | ○ |
| LR | ○ | ○ | ○ | ○ | ○ | ○ | ○ | ○ |
| Symmetry | $A_1$ | $A_1$ | $A_2$ | $A_2$ | $A_1$ | $A_1$ | $A_1$ | $A_1$ |

**Table II.** $|\langle \epsilon_o | R_j | \epsilon_i \rangle|^2$ corresponding to the vibration modes of $A_1$ and $A_2$.

| | HH | HV | LL | LR |
|---|---|---|---|---|
| $A_1$ | $a^2$ | $0$ | $\left\| \dfrac{a+b}{2} \right\|^2$ | $\left\| \dfrac{a-b}{2} \right\|^2$ |
| $A_2$ | $0$ | $d^2$ | $0$ | $d^2$ |

**Table III.** $|\langle \epsilon_o | R_j | \epsilon_i \rangle|^2$ in angle-resolved Raman spectroscopy.

| Symmetry | HH | HV |
|----------|----|----|
| $A_1$ | $|a + (-a + b)\sin^2\theta|^2$ | $\dfrac{|-a + b|^2}{4}\sin^2 2\theta$ |
| $A_2$ | $d^2\sin^2 2\theta$ | $d^2\cos^2 2\theta$ |

**Table IV.** Experimental Raman peak positions (our data) and calculated ones (Kong's group). Our data were obtained through a Lorentzian fit of the LR spectrum shown in Figure 3(b).

| Raman peak | P0 | P1 | P2 | P3 | P4 | P5 | P6 | P7 |
|---|---|---|---|---|---|---|---|---|
| Our experimental peak position ($cm^{-1}$) | 8.6 | 80.3 | 91.3 | 112.1 | 117.3 | 133.2 | 163.0 | 211.5 |
| Calculated peak position[20] ($cm^{-1}$) | 8.9 | 75.7 | 89.1 | 112.1 | 115.2 | 132.0 | 165.7 | 211.3 |
| Symmetry[20] (Caculation) | $A_1$ | $A_1$ | $A_2$ | $A_2$ | $A_1$ | $A_1$ | $A_1$ | $A_1$ |

**Table A·1.** The obtained fitting parameters for the experimental data shown in Figure A·1(a) and (b).

|  | $\left|\dfrac{b}{a}\right|$ | $\varphi$ |
|---|---|---|
| HH | 0.517 | $0.406\pi$ |
| HV | 0.405 | $0.387\pi$ |